Target Considerations for a Very Cold Neutron Source

By

B. J. Micklich

IPNS, Argonne National Laboratory, Argonne, IL

# TARGET CONSIDERATIONS FOR A VERY COLD NEUTRON SOURCE*

B. J. Micklich#, Argonne National Laboratory, Argonne, IL 60439, U.S.A.

*Abstract*

A proposed Very Cold Neutron Source (VCNS) would operate with 4 ms long pulses at 5 Hz and 1 GeV. The energy per pulse would be 300 kJ, much higher than the Spallation Neutron Source (33 kJ/pulse) or the present IPNS (0.3 kJ/pulse), and the peak power on target would be 75 MW. This paper discusses ideas for the VCNS target and examines the possibility of conducting target tests at an 8-GeV proton linear accelerator which is being studied for construction at Fermilab.

## A VERY COLD NEUTRON SOURCE

Many interesting scientific problems involve phenomena at longer length scales and slower time scales than can be addressed through neutron scattering experiments that use present-day neutron sources and instrumentation [1]. Many of the features that make neutrons useful probes (optical methods, phase shifts, gravitational and magnetic effects, high absorption in selected isotopes) are better exploited with longer-wavelength very-cold neutrons than with conventional thermal or cold neutrons. Neutrons can be focused and guided more efficiently at longer wavelengths (to increase intensity on the sample), making a source of very-cold neutrons desirable for experiments involving extreme environments or smaller or more weakly scattering samples. Longer wavelength neutrons could also be used to image living biological systems. Longer-wavelength, lower-energy neutrons would also be useful for experiments in fundamental physics such as measurement of the neutron electric dipole moment and half-life.

Table 1: Neutron source definitions

|  | energy (meV) | temperature (K) | wavelength (Å) |
|---|---|---|---|
| Cold | 5 | 60 | 4 |
| Very Cold | 0.2 | 2 | 20 |
| Ultra Cold | 0.0002 | - | 600 |

The motivation for studying a Very Cold Neutron Source (VCNS) is to establish the prospects for a neutron source which would provide intense pulsed beams with spectra "as cold as is realistic" (i.e., the longest practical wavelength distribution). Present cold neutron sources are designed to have the peak wavelength to be in the range of 2-4 Å. The desired peak wavelength for a VCNS is in the range of 10-20 Å, with usable flux extending out to 100 Å, in quantities several orders of magnitude greater than available from present or near-term envisioned sources. However, the source should still have significant intensity at shorter wavelengths to support traditional diffraction experiments as well as to provide a broad range of momentum and energy transfer [1]. The vision is that of a facility providing unique neutron scattering capabilities to the international community toward the end of the next decade, complementary to those available at facilities such as the Spallation Neutron Source (SNS).

A workshop to explore the scientific interest in a VCNS was held at Argonne on 21-24 August 2005 [2]. Thirty-nine participants from twelve US and international institutions attended the workshop. The primary purpose was to gauge interest in scientific applications of a facility delivering 30-1000 times the currently available flux at 20 Å. The reports of the workshop discussion groups (in scientific applications, instruments and techniques, and sources) influenced considerably the facility concept described in this paper.

*Accelerator*

Preliminary neutronics studies of potential VCNS moderators indicated that the resulting neutron emission time distributions would have widths on the order of a few milliseconds for the moderators and neutron wavelengths of interest. It is then suitable to use a long-pulse source with the accelerator pulse length chosen to not perturb greatly the natural pulse width from the moderator. Since pulse compression is not required, a linear accelerator can provide the full beam energy. The use of long-wavelength neutrons implies a long interpulse interval, and thus a low pulsing frequency.

Using 1000 MeV protons in pulses which are 4 ms long with a peak current of 75 mA, the energy per pulse will be 300 kJ. The likely scientific applications could profitably use five pulses/sec [2], so that the average proton beam power would be 1500 kW. Figure 1 shows one feasible accelerator structure for the VCNS linac. Any required accelerator development can be done in collaboration with the high-energy physics community, as similar structures are being proposed for superconducting proton and electron linear accelerators [3,4].

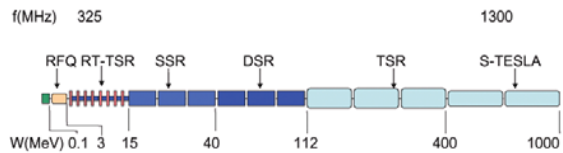

Figure 1. Schematic diagram of VCNS linear accelerator. Structures and corresponding energy ranges. RFQ: Radio-Frequency Quadrupole; RT-TSR: Room Temperature Triple Spoke Resonator; SSR: (Superconducting) Single Spoke Resonator; DSR: (Superconducting) Double Spoke Resonator; TSR: (Superconducting) Triple Spoke Resonator; S-TESLA: Superconducting TESLA-type elliptical cavity.

___________________________________________
*Work supported by US Department of Energy, Office of Science.
#bjmicklich@anl.gov

*Target*

The target should be composed of a high-Z material capable of high-power operation and also exhibiting a low cross section for neutron absorption (see Figure 2). Of the candidate materials, bismuth has the lowest neutron absorption but its use leads to production of $^{210}$Po, potentially a significant radiation hazard. Mercury exhibits substantial neutron absorption. Lead is another high-Z material that could operate comfortably at this power level. Neutron absorption in lead, though higher than in bismuth, is still small and is not expected to lead to excessively large neutron capture losses. One intriguing possibility is the use of radiogenic lead, which is low in the neutron absorbing isotopes $^{204}$Pb and $^{207}$Pb. Lead-bismuth eutectic (LBE) could also be used if the concerns about $^{210}$Po are not too great, as it has a significantly lower melting point than lead (125 ºC vs. 327 ºC) and its use would lead to some operational simplifications.

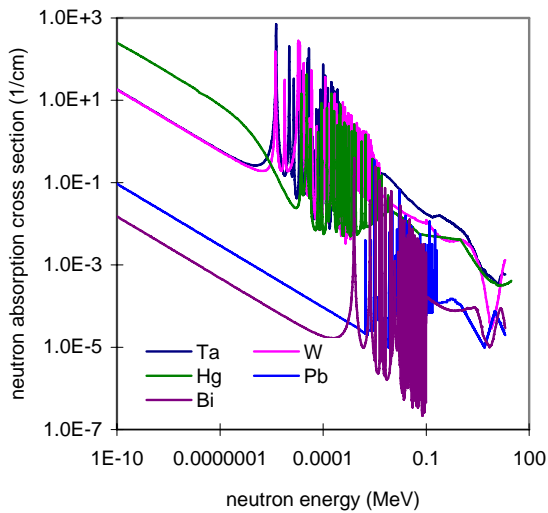

Figure 2. Neutron absorption cross sections for candidate target materials.

The original VCNS concept had a single 300 kJ pulse per second, for an average power of 300 kW. The Canadian TRIUMF Thermal Neutron Facility [5] operated with an edge-cooled solid target at a power level of 200 kW and that design could be extended to the VCNS. However, at a pulse repetition rate of 5 Hz and the same number of protons/pulse, the target could probably not be edge cooled, since the average power would then be 1.5 MW. In this case the target might have to be of a traditional plate design, with coolant flow between the plates, or like the "cannelloni" target [6] being built for the ultra-cold neutron source at PSI. Alternatively, an externally-cooled circulating target of lead, bismuth, or LBE might serve the same purpose. The technology is well established for all three types of targets. Advantage can be taken of the fact that the world community is accumulating a significant operating experience with high-power spallation targets at SINQ, at the SNS, and at JSNS.

The ability to cool the target entrance window and the high-energy portion of the target while supporting stresses in the target structure will place a maximum limit on the allowable current density at the target window. Based on a set of earlier simulations [7], we have chosen a target diameter of 20 cm and a proton beam diameter of 15 cm. For this case the time-averaged current density at the front window is about 8.5 µA/cm$^2$, and the average beam power on target would be about 8.5 kW/cm$^2$, about the same as that for the SINQ target [8]. The large target diameter (20 cm) will help to avoid problems with excessively high volumetric heating in the target.

Even though VCNS would have 10x the energy per pulse as SNS, cavitation problems are not expected for a liquid target in long-pulse operation. Calculations show that only first few hundred microseconds of the 4-ms pulse would contribute to development of the pressure pulse in the target. Turning on the accelerator pulse over a 200-µs time period might mitigate some of the effects.

*Moderator*

In the context of VCNS, the "lowest practical temperature" means the use of liquid helium as the moderator coolant, taking advantage of the properties (high heat capacity and thermal conductivity) of superfluid LHe below the λ point (2.17 K). The moderator would be in the form of solid pellets (such as those shown in Figure 3) [9] cooled by superfluid helium flowing in the spaces between the pellets. By manipulating the pellet shape and packing, we can 'tune' the relative volume of moderator and coolant in order to adjust the moderating power or the cooling capacity of the moderator/coolant system. Production of large quantities of mm-scale, solid pellets (e.g., $CO_2$, $CH_4$, $NH_3$, $D_2$,) is an established technology [9]. The same technology should be applicable to $D_2O$, which would produce low-density amorphous (LDA) ice.

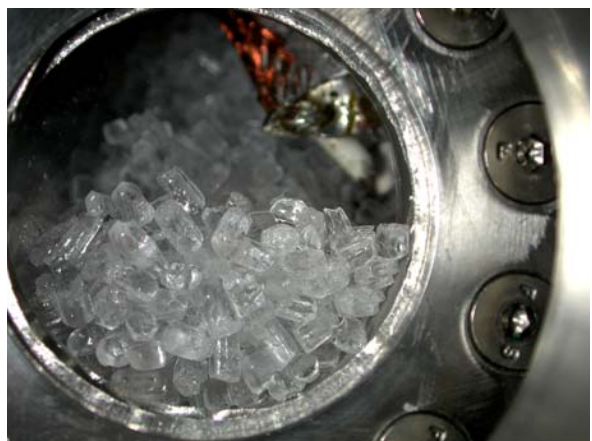

Figure 3. Cryogenic methane moderator pellets (on the order of a few mm in size) [9].

The design and technology of a VCNS will be very similar to those of UCN sources. Both types of sources will use a warm moderator in combination with a cold or ultra-cold moderator. For the warm moderator, $D_2O$ is preferred over $H_2O$ because of its much lower neutron absorption. The very cold moderator would also rely on deuterated compounds (e.g., solid phases of $D_2$, $D_2O$, $CD_4$, etc.) rather than their hydrogenated counterparts. The moderator material should exhibit incoherent scattering at low temperatures and low neutron energies. The Bragg edges in the scattering cross sections of materials such as beryllium and graphite make these poor choices since below the lowest Bragg edge the scattering cross section is essentially zero.

Moderator heating (both the total energy deposition in the moderator and the peak energy deposition in space and time) is a new and critical aspect of a VCNS. The instantaneous heating of the moderating medium (the VCNS accelerator would deliver 300 kJ/pulse vs. 0.34 kJ/pulse for IPNS or 33 kJ/pulse for SNS) will raise the moderator temperature during the pulse. Heating must be low enough to keep the short-term temperature rise within acceptable limits, and the cooling capacity must be sufficient to maintain the low time-average temperature. The fundamental constraints are the low heat transport rates and low heat capacities of cryogenic materials. Unfortunately, data on low-temperature thermal conductivities and specific heats of candidate moderator materials are sparse, yet are essential to evaluating the static and dynamic effects of moderator heating.

Preliminary heat transfer calculations in the liquid helium show that the heat conduction through the helium may be a greater limitation than either thermal diffusion within the pellets or the Kapitza resistance at the pellet surface [10]. Energy deposition may be limited by heat conduction to about 50-100 mW/ $cm^3$.

## NEUTRON SOURCE CONSIDERATIONS

### Fermilab Proton Accelerator

Protons are provided to the Main Injector at Fermilab by a 400-MeV linear accelerator and an 8-Gev booster synchrotron. In order to increase the proton current available for injection into the Main Ring, Fermilab had proposed the construction of an 8-GeV linear accelerator called the Proton Driver [3]. More recently the parameters of the accelerator have been altered to emphasize the applications to an electron collider. The new accelerator, termed Project X [4], has lower performance for protons in terms of peak current and energy per pulse than the Proton Driver. Because of the physical proximity of the Argonne and Fermi laboratories, their common management by the University of Chicago, the natural collaboration between the laboratories on other accelerator issues, and the similarity in accelerator performance parameters between VCNS, the Proton Driver, and Project X, it is interesting to examine the uses of the proposed 8-Gev accelerator for neutron production.

### Comparison of Proton Linac Parameters

Table 2 shows the main performance parameters for the VCNS accelerator concept described above, the ultimate version of the Proton Driver, and the Project X accelerator. The energy per pulse of the Proton Driver was consistent with that for the VCNS since the lower peak current and shorter pulse length were compensated by the higher beam energy. However, the energy per pulse of the Project X linac is only about one-quarter that desired for the VCNS because its peak current is further reduced.

Table 2: Parameters for VCNS, Proton Driver, and Project X accelerators.

| Parameter | VCNS | Proton Driver | Project X |
|---|---|---|---|
| $E_p$ (GeV) | 1 | 8 | 8 |
| pulse length (ms) | 4 | 1 | 1 |
| rep. rate (1/s) | 5 | 10 | 5 |
| avg. current (mA) | 1.5 | 0.25 | 0.045 |
| peak current (mA) | 75 | 25 | 9 |
| avg. power (MW) | 1.5 | 2.0 | 0.36 |
| energy/pulse (kJ) | 300 | 200 | 72 |

The physics program proposed for the Project X proton accelerator requires five pulses per second at the level of performance indicated in Table 2. Should that accelerator also serve to drive a VCNS, an additional five pulses per second could be interleaved with them, at a higher peak current or longer pulse length (e.g., 3-4 ms) in order to deliver pulses with higher energy per pulse to a VCNS spallation target, as long as the necessary RF power were available. The additional klystrons and higher-power RF couplers should be installed during the original facility construction. While they could be added later, that approach would add considerably to the overall facility costs.

### Optimal Energy for Spallation Neutron Sources

There has been much discussion about the optimal energy for spallation neutron sources. The major points are well summarized by Carpenter et al. [11]. The issue goes beyond the energy at which neutron production is the most efficient, as technical and engineering issues are extremely important for high-power accelerator targets. It is easier to approach multi-megawatt beam power by going to high energy rather than high current. Lower current reduces problems due to space charge limits at the low-energy end of the accelerator, and higher energy reduces energy deposition in the target window and the front of the target. On the other hand, lower energies lead to a more compact source (and thus potentially higher fluxes from the moderator), reduced shielding requirements, and reduced costs for a linear accelerator.

Figure 4 illustrates the issue of neutron production efficiency. At higher proton energies, an increasing fraction of the incident beam power is diverted into non-hadronic channels, reducing the overall neutron production efficiency per unit input power. Table 3 shows these results in a different format. For each proton energy, the hadronic energy fraction $F_h$ gives the fraction of incident power that goes into neutron production. Table 3 also compares the proton current needed for a constant power of 1 MW ($I_1$) with the current needed for a constant neutron source rate ($I_n$) at the level of a 1-MW, 1 GeV proton beam. The 8-GeV Project X linac would have to supply an extra 20% power to maintain the same neutron source rate at the baseline VCNS linac.

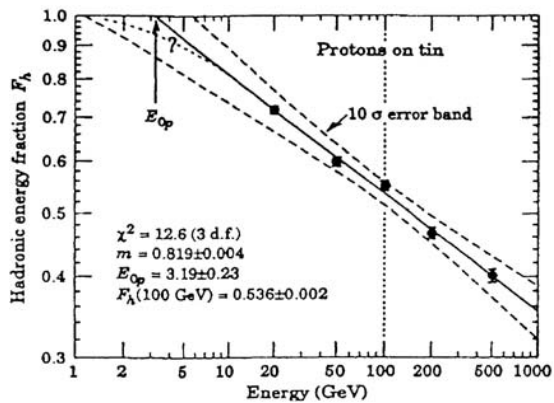

Figure 4. Hadronic energy fraction in shower due to incident protons vs. incident proton energy [11].

## Neutron Source Calculations

MCNPX [12] was used to calculate the neutron source distributions in lead targets for selected incident proton energies between 0.75 and 10 GeV and an incident beam power of 1 MW. Figure 5 shows the results. In the figure, the intensity color scale is the same for all proton energies. As the incident energy increases, the peak volumetric neutron source decreases and also moves further into the target. Figure 6 shows the corresponding neutron emission from the sides of the target. Notice that even for cases where the maximum neutron production is at the front of the target, the peak neutron emission occurs somewhat downstream due to the forward-directed emission angle of neutrons from the source reactions. The axial position at which the neutron emission is a maximum increases with incident proton energy.

Table 3: Variation of accelerator current with energy for constant neutron source strength.

| $E_p$ (GeV) | $F_h$ | $I_1$ (mA) | $I_n$ (mA) | $I_n/I_1$ |
|---|---|---|---|---|
| 1 | 1.0 | 1.0 | 1.0 | 1.0 |
| 2 | 0.97 | 0.5 | 0.515 | 1.03 |
| 3 | 0.94 | 0.333 | 0.353 | 1.06 |
| 5 | 0.89 | 0.2 | 0.224 | 1.12 |
| 8 | 0.84 | 0.125 | 0.149 | 1.19 |
| 10 | 0.815 | 0.1 | 0.123 | 1.23 |
| 20 | 0.72 | 0.05 | 0.0695 | 1.39 |

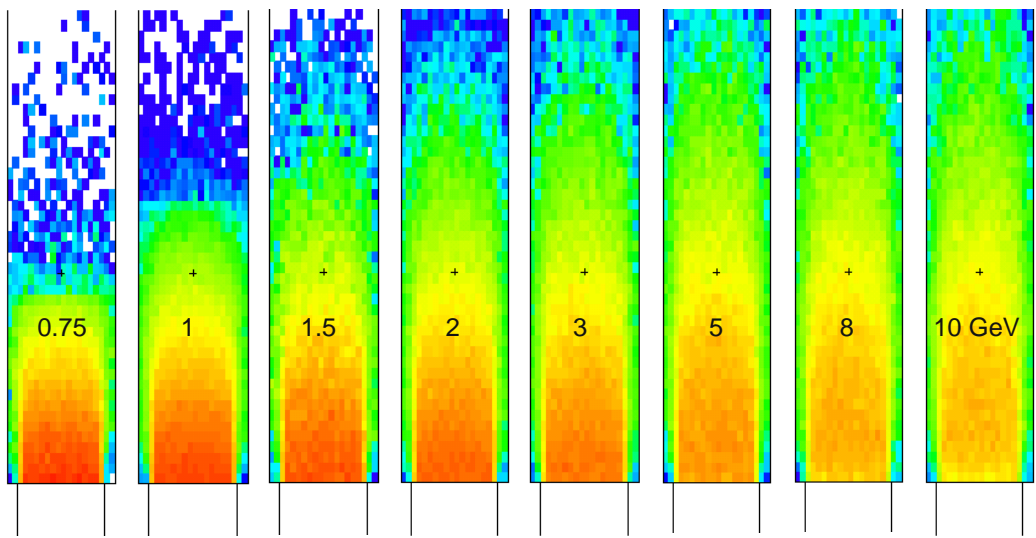

Figure 5. MCNPX mesh tally plots of neutron generation in lead targets for selected incident proton energies between 0.75 and 10 GeV. A parallel proton beam of 15 cm diameter is incident from the bottom on a 20-cm diameter target. The intensity color scale is the same for all plots.

Figure 7 shows the neutron yield per proton, divided by the incident proton beam energy, for the same simulations. Presented in this way, the results indicate the most efficient proton beam energy for neutron generation. The data show that neutron production is most efficient at about $E_p = 1.5$ GeV, falling off by almost 20% at 8 GeV. This is consistent with the data shown in Table 3.

Neutron source plots for 1-GeV protons incident on lead, tungsten, and mercury targets are shown in Figure 8. The peak neutron production occurs nearer the front of the target for the denser materials. Figure 9 shows the corresponding neutron emission from the sides of the target.

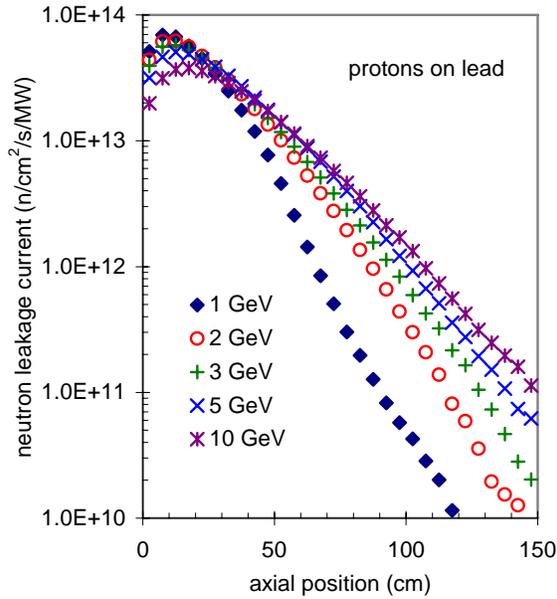

Figure 6. Axial dependence of neutron leakage from the side of a lead spallation target for selected proton energies (1 MW incident proton beam).

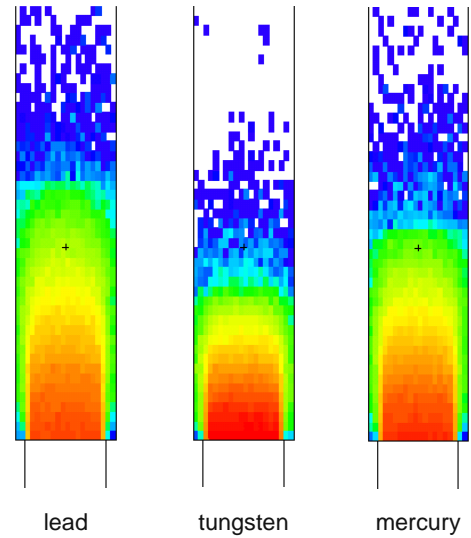

Figure 8. MCNPX mesh tally plots of neutron generation in targets of selected materials under irradiation with 1-GeV protons. A parallel proton beam of 15 cm diameter is incident from the bottom on 20-cm diameter targets. The intensity color scale is the same for all plots.

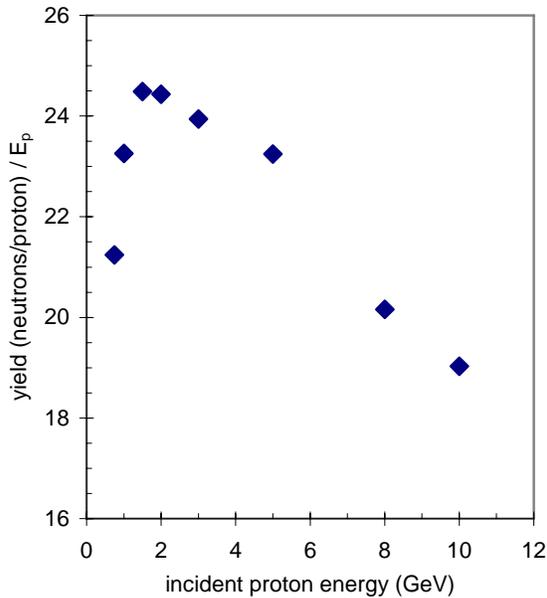

Figure 7. Neutron yield per proton divided by incident proton energy as a function of proton energy.

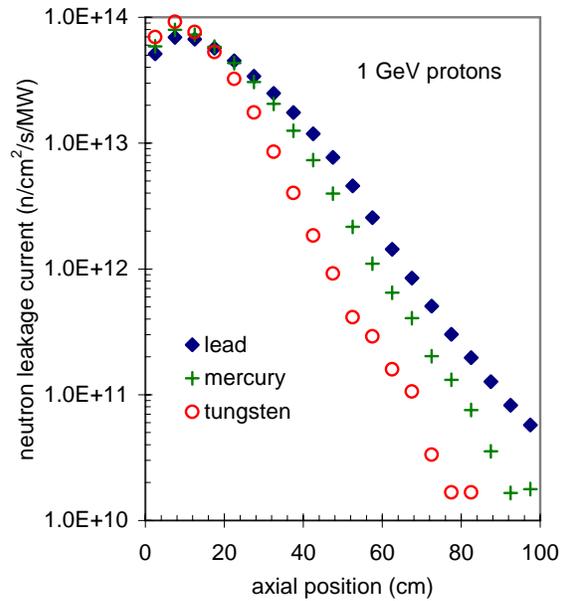

Figure 9. Axial dependence of neutron leakage from the side of spallation targets due to irradiation with 1-GeV protons.

Figure 10 shows the MCNPX results for power density in lead targets for selected proton energies. As the proton energy increases, the maximum power density decreases and moves into the target rather than occurring just behind the target front face. The maximum power density for 10 GeV incident protons is only about one-half of that for 1 GeV protons, which should make the technical problems of target cooling somewhat easier and might even make possible the use of a solid target rather than a liquid target for incident beam power greater than 1 MW. However, detailed thermal hydraulics calculations are required to determine whether the physical limits on internal temperature and surface heat flux for a plate-type solid target can be met without diluting the beam interaction region with coolant to the point that the neutron production is diminished.

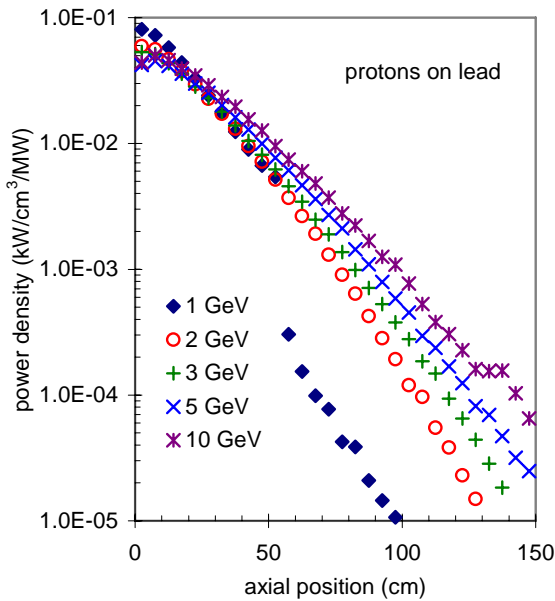

Figure 10. Axial variation of power density in lead targets irradiated by protons of selected energies.

## Neutron Flux Calculations

In the next set of calculations, a lead target was placed in the center of a liquid $D_2O$ moderator/reflector (see model in Figure 11). The thermal neutron flux was calculated as a function of radial and axial position and is shown in Figure 12 for selected incident proton energies. As the proton energy increases, the maximum thermal neutron flux decreases and moves further back into the target. However, the neutron flux at about 20 cm radially from the target remains roughly constant. Thus a cold moderator placed in this location would deliver about the same cold neutron flux for each incident proton energy.

## Neutron Emission Calculations

The results of the neutron flux calculations in the warm moderator were used to determine the cold neutron emission for 1-MW proton beams with energies of 1 and 8 GeV. For each proton energy the cold moderator was placed at an axial position corresponding to the axial position of the thermal flux maximum shown in Figure 12. The moderator center was 13.5 cm behind the front of the target for 1 GeV incident protons, and 23.5 cm behind the target front for 8 GeV incident protons. The MCNPX model for the 1 GeV proton case is shown in Figure 13. The cold moderator in these calculations was a 20 cm x 20 cm x 10 cm thick rectangular region of liquid $D_2$ at 20 K, enclosed in a magnesium vacuum jacket.

The scattering kernel for liquid $D_2$ is the lowest temperature scattering kernel available for the candidate materials, and was used in these calculations to estimate the dependence of long-wavelength neutron flux on premoderator thickness. Realistic neutronic simulations for a VCNS require scattering kernels that are largely unavailable for the materials and temperatures of interest. Developing these kernels requires considerable effort and a detailed knowledge which is possessed by only a few experienced practitioners. The existing models that describe the physics of neutron scattering are good for some materials (graphite, beryllium) but need development for others (deuterium, heavy water, deuterated methane).

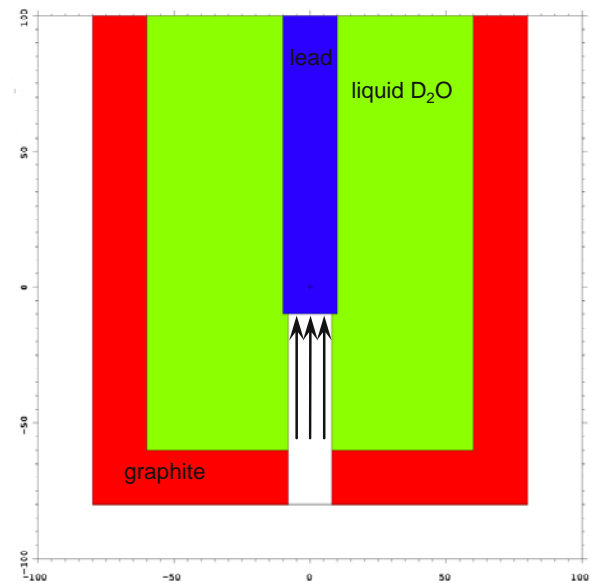

Figure 11. MCNPX model used to calculate thermal neutron flux in the liquid $D_2O$ moderator/reflector.

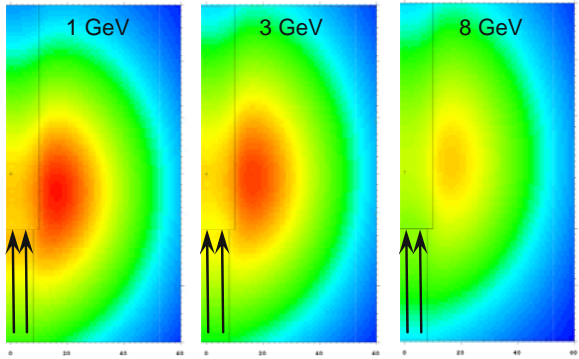

Figure 12. Mesh tally plots of thermal neutron flux in a D$_2$O moderator surrounding a lead accelerator target. The intensity color scale is the same for all plots.

The main problem to be overcome in the VCNS design is to reduce the nuclear heating (neutron and gamma) in the cold moderator while maintaining a high neutron flux. Because nuclear heating (fast neutrons and gammas from the source) decreases exponentially with distance from the source region, while neutrons, in the absence of absorption, are preserved while slowing down, we expect to find a large radius at which the heating is acceptable and the cold neutron flux is as large as possible. The source power in this approach is an adjustable parameter, which is a departure from convention in neutron facility design. The parameter to be optimized is the ratio of cold neutron flux to nuclear heating in the moderator. This ratio improves with distance from the source, but at the sacrifice of the ratio of cold neutron flux to source power.

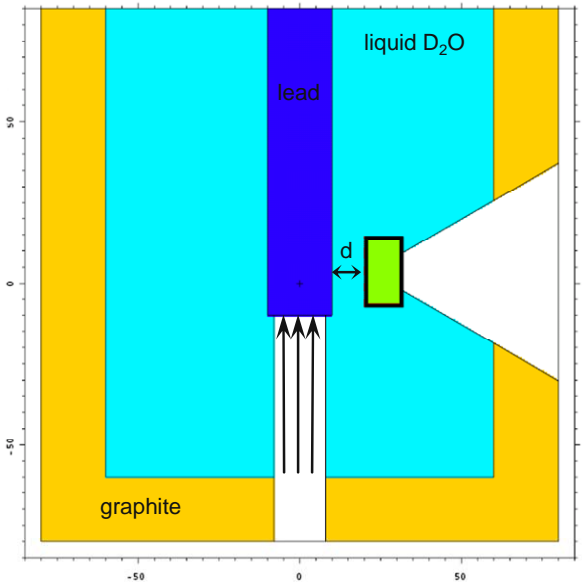

Figure 13. MCNPX model used to calculate cold neutron emission from a liquid D$_2$ moderator. The premoderator thickness (d) was varied in the simulations.

This physical argument explains the phenomena shown in Figure 15. As the premoderator thickness increases, both the total energy deposition and the peak volumetric deposition decrease rapidly. The peak heating falls below the value 100 mW/cm$^3$ when the premoderator thickness is about 17 cm. Figure 15 shows that, while the total long-wavelength flux decreases slowly as the moderator is moved farther from the target, the long-wavelength flux per energy deposited in the moderator increases rapidly. Since the goal is to maximize long-wavelength flux, and the primary constraint is maximum local energy deposition in the moderator, this means that the moderator should be located far from the neutron producing target. If the accelerator power were reduced by 50%, the moderator could be moved closer to the target and remain under the 100 mW/cm$^3$ limit, but would actually yield a lower long-wavelength neutron flux.

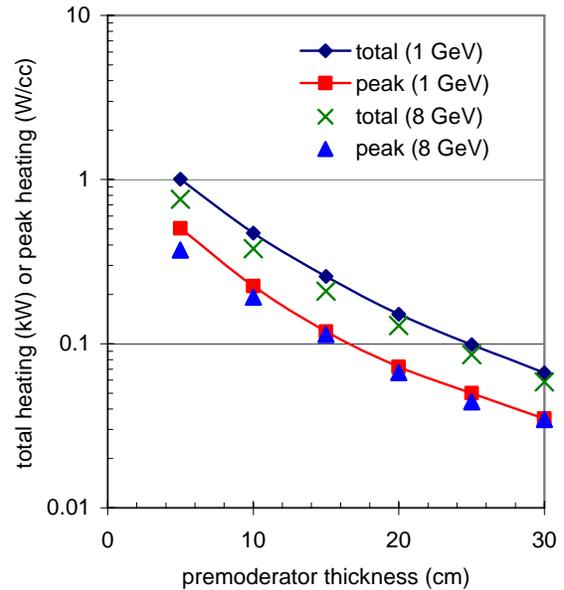

Figure 14. Total heating (kW) and peak heating (mW/cc) as functions of premoderator thickness. Results are for the computational model shown in Figure 13.

## CONCLUSIONS

A Very Cold Neutron Source would be designed to provide the maximum flux of neutrons having wavelengths greater than 10 Å. While the optimum proton beam energy is about 1.5 GeV from the point of view of neutron production efficiency, some considerations (target window heating, current limits at the low-energy end of the accelerator, and beam losses causing activation in the accelerator) would favor higher beam energies. While lower incident proton energies are somewhat better for producing long-wavelength neutrons (because of the lower average neutron source energy), the long wavelength neutron flux normalized to the maximum volumetric energy deposition in the cold moderator appears to be independent of proton energy. Solid targets may still be used within technological limits. Proton energies up to 10 GeV are suitable for a VCNS linac, and useful information would be obtained by performing target tests on accelerators of this energy.

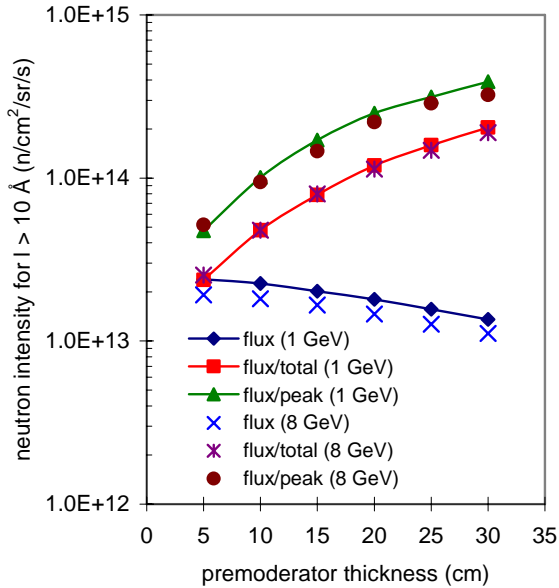

Figure 15. Long-wavelength flux, flux per total heating, and flux per maximum heating as functions of premoderator thickness. Results are for the computational model shown in Figure 13.